\pdfoutput=1
\documentclass[conference]{IEEEtran}
\IEEEoverridecommandlockouts
\usepackage{cite}
\usepackage{amsmath,amssymb,amsfonts}
\usepackage{graphicx}
\usepackage{xcolor}
\usepackage{booktabs}
\usepackage{url}
\usepackage{textcomp}
\usepackage{enumitem}
\usepackage{tikz}
\makeatletter
\newcommand{\linebreakand}{%
  \end{@IEEEauthorhalign}
  \hfill\mbox{}\par
  \mbox{}\hfill\begin{@IEEEauthorhalign}}
\makeatother
\usepackage{eso-pic}
\AddToShipoutPictureBG*{\AtPageLowerLeft{\put(\LenToUnit{0.5\paperwidth},\LenToUnit{0.30in}){\makebox[0pt]{\parbox{0.86\paperwidth}{\centering\scriptsize
\copyright~2026 IEEE. Personal use of this material is permitted. Permission from IEEE must be obtained for all other uses, in any current or future media, including reprinting/republishing this material for advertising or promotional purposes, creating new collective works, for resale or redistribution to servers or lists, or reuse of any copyrighted component of this work in other works.}}}}}
\usetikzlibrary{positioning,arrows.meta,fit,backgrounds,shapes.geometric}

\graphicspath{{figures/}}

\def\BibTeX{{\rm B\kern-.05em{\sc i\kern-.025em b}\kern-.08em
    T\kern-.1667em\lower.7ex\hbox{E}\kern-.125emX}}
\begin{document}
\bstctlcite{BSTcontrol}

\title{SnapScope: A Platform for City-Scale Collection and Exploration of Public Snap Map Data}

\author{\IEEEauthorblockN{Mohammed Almukaynizi, \IEEEmembership{Member, IEEE}}
\IEEEauthorblockA{\textit{Dept. of Information Systems} \\
\textit{College of Computer and}\\
\textit{Information Sciences}\\
\textit{King Saud University}\\
Riyadh, Saudi Arabia \\
malmukaynizi@ksu.edu.sa \\
https://orcid.org/0000-0001-9553-0896}
\and
\IEEEauthorblockN{Fahad Alhaqbani}
\IEEEauthorblockA{\textit{Dept. of Information Systems} \\
\textit{College of Computer and}\\
\textit{Information Sciences}\\
\textit{King Saud University}\\
Riyadh, Saudi Arabia \\
443102246@student.ksu.edu.sa}
\and
\IEEEauthorblockN{Khaled Almarzoug}
\IEEEauthorblockA{\textit{Dept. of Information Systems} \\
\textit{College of Computer and}\\
\textit{Information Sciences}\\
\textit{King Saud University}\\
Riyadh, Saudi Arabia \\
443101266@student.ksu.edu.sa}
\linebreakand
\IEEEauthorblockN{Sultan Alanbari}
\IEEEauthorblockA{\textit{Dept. of Information Systems} \\
\textit{College of Computer and}\\
\textit{Information Sciences}\\
\textit{King Saud University}\\
Riyadh, Saudi Arabia \\
443101437@student.ksu.edu.sa}}

\maketitle

\begin{abstract}
Snapchat's Snap Map is an ephemeral stream of geotagged public video and
image stories. The platform documents no API for it, no prior work
describes a reproducible way to collect it at city scale, and no tool
exists for exploring what is collected. We present SnapScope, which
pairs a collection back end with a web front end. The back end builds a
geographic query grid, sweeps it on a schedule, and deduplicates at
ingestion; the front end supports scraper management, interactive
exploration, neighborhood comparison, and export. Deployed over Riyadh,
Saudi Arabia, it collected 515{,}364 unique public snaps posted across
23~days (17~September--9~October~2024) on a 1\,km grid of
2{,}740~query points. A saturation probe over 21~consecutive runs found
94.8\% of returned observations already stored, indicating that
successive passes mostly return already-captured content. We release a
privacy-safe aggregate dataset of tile-level hourly counts under
CC~BY~4.0. The platform is city-agnostic: redeployment requires only
new grid coordinates and boundary polygons.
\end{abstract}


\section{Introduction}\label{sec:intro}

Geotagged social media has become a widely used proxy for urban activity.
Posts, check-ins, and shared stories from Twitter, Flickr, and Foursquare
have been mined to map points of interest~\cite{crandall2009}, reveal
mobility patterns~\cite{noulas2011}, delineate functional
neighborhoods~\cite{cranshaw2012}, and detect events in real
time~\cite{sakaki2010}. Yet nearly all of this work relies on platforms
whose content is textual, persistent, and accessible through documented
APIs~\cite{goodchild2007}.

Snapchat's Snap Map operates under a different model. Public stories posted to the
Map are visual (short videos or images), hyper-local (pinned to
coordinates), and ephemeral---a design choice that shapes user behavior
toward spontaneous sharing~\cite{bayer2016}. That
combination gives Snap Map near-real-time coverage of local urban
activity, but the same ephemerality that makes the data timely also makes
systematic collection hard. The platform has been used as an open-source intelligence tool
for monitoring social unrest~\cite{matthews2021}. Snapchat provides no official API, and the
handful of quantitative studies that
do exist~\cite{juhasz2018,juhasz2019,alghamdi2019b} offer no
reproducible collection methodology and no tooling for managing or
exploring what gets collected.

We present SnapScope, an integrated platform with two layers:
\begin{enumerate}[nosep]
  \item A \emph{collection back end} that uses undocumented Snap Map
    API endpoints (previously identified in prior work and public
    repositories) within an automated pipeline: geographic grid
    construction, scheduled multi-pass collection with ingestion-time
    deduplication, and coverage saturation measurement.
  \item A \emph{web-based front end} that provides scraper management
    (create, monitor, stop, update, delete), data filtering by district
    and time, side-by-side neighborhood comparison with interactive
    charts, role-based user management, and data export.
\end{enumerate}
We deploy the platform over Riyadh, Saudi Arabia, where Snapchat
is widely adopted. Nationwide, Snapchat's potential advertising reach
covers 91.8\% of the Saudi population aged 13 and
above~\cite{datareportal2026}. The deployment produced a dataset of
515{,}364~snaps that we characterize at the dataset level and release
in privacy-safe aggregate form. The contribution here is the \emph{platform and dataset}. Spatial
and temporal analyses of the collected data are presented separately.

\section{Related Work}\label{sec:related}

The citizens-as-sensors paradigm~\cite{goodchild2007} established
volunteered geographic information (VGI) as a complement to
authoritative spatial data. The research that followed is well
established: mapping from geotagged photos~\cite{crandall2009},
extracting mobility from check-ins~\cite{noulas2011,cheng2011},
detecting events from tweets~\cite{sakaki2010,weng2011}, inferring
land use from location-based social network activity~\cite{frias2014},
and the broader urban computing landscape~\cite{zheng2014}. All of it
relies on platforms with documented APIs and persistent content.
Ethical and quality concerns around VGI collection are well
documented~\cite{crampton2013,elwood2012}, and even coarse mobility
traces can re-identify individuals~\cite{demontjoye2013}---motivating
the privacy-by-design choices in our platform.

Snap Map has attracted far less scholarly attention, and what exists
focuses on spatial characterization or event monitoring rather than on
collection infrastructure: Juh\'{a}sz and
Hochmair~\cite{juhasz2018,juhasz2019} characterized spatial and
temporal dynamics in U.S.\ cities and Florida state parks, Alghamdi
et~al.~\cite{alghamdi2019b} analyzed crowd behavior at the Grand Mosque
in Makkah during Ramadan, and Matthews et~al.~\cite{matthews2021}
demonstrated its use as an open-source surveillance tool. None
documents a reusable collection system, provides management or
exploration tooling, discusses coverage saturation, or releases a
dataset. Our work addresses all four gaps.

\section{System Architecture}\label{sec:system}

SnapScope is organized into two layers: a collection back end that
interfaces with the Snap Map API and a web front end that lets
operators and researchers interact with the results
(Fig.~\ref{fig:architecture}). The layers communicate through a
RESTful API served by FastAPI.

\begin{figure*}[!t]
  \centering
  \resizebox{\textwidth}{!}{%
  \begin{tikzpicture}[
    node distance=0.6cm and 1.0cm,
    box/.style={draw, rounded corners=3pt, minimum height=1.1cm,
                minimum width=2.6cm, align=center, font=\small\sffamily},
    smallbox/.style={draw, rounded corners=2pt, minimum height=0.8cm,
                minimum width=2.0cm, align=center, font=\scriptsize\sffamily},
    db/.style={draw, cylinder, shape border rotate=90, aspect=0.25,
               minimum height=1.1cm, minimum width=2.0cm,
               align=center, font=\small\sffamily},
    arr/.style={-{Stealth[length=5pt]}, thick},
    lbl/.style={font=\scriptsize\sffamily, midway, above, text=black!70},
    lblbelow/.style={font=\scriptsize\sffamily, midway, below, text=black!70},
  ]
    \node[box, fill=orange!15] (snapmap) {Snap Map\\Platform};

    \node[box, fill=blue!12, right=1.2cm of snapmap] (tileset) {getLatestTileSet\\{\scriptsize(epoch ID)}};
    \node[box, fill=blue!12, right=of tileset] (playlist) {getPlaylist\\{\scriptsize(per grid point)}};
    \node[box, fill=blue!12, right=of playlist] (dedup) {Deduplication\\{\scriptsize(media ID check)}};

    \node[db, fill=green!12, right=1.6cm of dedup] (mongo) {MongoDB\\{\scriptsize(snap metadata)}};
    \node[db, fill=yellow!15, below=0.8cm of mongo] (pg) {PostgreSQL\\{\scriptsize(users, config)}};

    \node[box, fill=blue!12, right=1.0cm of mongo] (fast) {FastAPI\\REST API\\{\scriptsize(JWT auth)}};

    \node[smallbox, fill=teal!12, right=1.2cm of fast, yshift=1.2cm] (scraper) {Scraper\\Management};
    \node[smallbox, fill=teal!12, right=1.2cm of fast, yshift=0.0cm] (dashboard) {Data Explorer\\+ Dashboard};
    \node[smallbox, fill=teal!12, right=1.2cm of fast, yshift=-1.2cm] (compare) {Neighborhood\\Comparison};

    \node[smallbox, fill=red!10, below=0.8cm of tileset] (grid) {1\,km Grid\\{\scriptsize(2,740 points)}};

    \node[smallbox, fill=purple!10, below=0.8cm of playlist] (sched) {Cron\\Scheduler};

    \node[smallbox, fill=gray!15, right=0.6cm of compare] (csv) {CSV\\Export};

    \draw[arr] (snapmap) -- node[lbl]{HTTPS} (tileset);
    \draw[arr] (tileset) -- node[lbl]{epoch} (playlist);
    \draw[arr] (playlist) -- node[lbl]{JSON} (dedup);
    \draw[arr] (dedup) -- node[lbl]{new snaps} (mongo);
    \draw[arr] (mongo) -- node[lbl]{read} (fast);
    \draw[arr] (fast) -- (scraper);
    \draw[arr] (fast) -- (dashboard);
    \draw[arr] (fast) -- (compare);
    \draw[arr] (dashboard) -- (csv);
    \draw[arr] (pg) -| (fast);
    \draw[arr] (fast) |- (pg);
    \draw[arr] (grid) -- (playlist);
    \draw[arr] (sched) -- (playlist);

    \begin{scope}[on background layer]
      \node[draw=blue!50, dashed, rounded corners=5pt, fill=blue!3,
            inner sep=8pt,
            fit=(tileset)(playlist)(dedup)(grid)(sched),
            label={[font=\small\sffamily, text=blue!70]above:Collection Engine (Python)}] {};
      \node[draw=green!50, dashed, rounded corners=5pt, fill=green!3,
            inner sep=8pt,
            fit=(mongo)(pg),
            label={[font=\small\sffamily, text=green!60!black]above:Storage Layer}] {};
      \node[draw=teal!50, dashed, rounded corners=5pt, fill=teal!3,
            inner sep=8pt,
            fit=(scraper)(dashboard)(compare)(csv),
            label={[font=\small\sffamily, text=teal!70]above:Front End (Next.js)}] {};
    \end{scope}
  \end{tikzpicture}}%
  \caption{SnapScope system architecture: collection engine, metadata store, REST layer, and web front end.}
  \label{fig:architecture}
\end{figure*}

\subsection{Collection Back End}\label{sec:backend}

\subsubsection{API Discovery and Validation}
Snap Map has no official API. By inspecting network traffic and publicly
available code repositories, we identified two HTTPS POST endpoints at
\texttt{ms.sc-jpl.com}: \texttt{getLatestTileSet}, which returns a
current epoch identifier for the map's data layer, and
\texttt{getPlaylist}, which accepts a geographic coordinate and returns
metadata for nearby public snaps. The key request parameters are
\texttt{requestGeoPoint} (query centroid) and \texttt{tileSetId.epoch}
(the current data snapshot, fetched via \texttt{getLatestTileSet}).
Additional fields---\texttt{radiusMeters}, \texttt{zoomLevel},
\texttt{maximumFuzzRadius}---showed no observable effect across dozens
of test queries.

To verify that publicly posted snaps actually appear through the API,
we uploaded 10~short test videos from three accounts on three
smartphones (Android and iOS) on 19--20~September~2024. All~10 appeared
on the public map within 12~minutes and were retrieved with matching
metadata. This round-trip test confirms retrievability and metadata
accuracy but does not measure what fraction of all eligible public snaps
the API chooses to surface---coverage completeness remains unknown
(Section~\ref{sec:limitations}).

\subsubsection{Grid Design}
\begin{figure}[!t]
  \centering
  \includegraphics[width=\columnwidth]{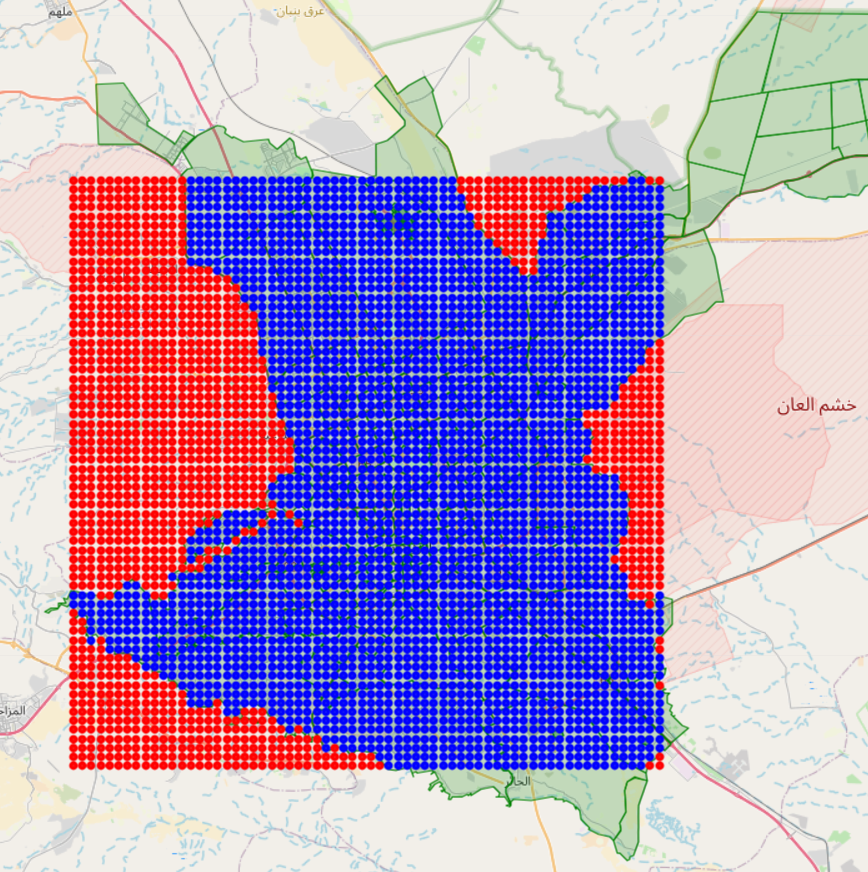}
  \caption{Query grid for Riyadh: 2{,}740 within-district centroids (blue); the 1{,}616 filtered points (red) are never queried.}
  \label{fig:grid}
\end{figure}

We built a square lattice of 4{,}356~points spaced 1\,km apart and
centered on Riyadh (24.71\textdegree N, 46.68\textdegree E), extending
to a 33\,km half-width. The 1\,km spacing was chosen to balance
coverage and cost: coarser grids risk missing localized clusters,
while finer grids multiply query volume with diminishing returns,
since the API already aggregates nearby content.
Points falling outside administrative district
boundaries~\cite{homaily} were filtered out, leaving 2{,}740~queried
centroids. Each scraper pass issues one API request per
point, dispatched through a bounded pool of ten concurrent workers.
Because each request covers a 1\,km radius on a 1\,km lattice, adjacent
query disks overlap substantially, and a snap near a cell boundary can
be returned at several neighboring points. Figure~\ref{fig:grid} shows the
resulting grid.

\subsubsection{Scheduling, Deduplication, and Storage}
The collection engine is a Python service that runs on a scheduled loop,
executing complete grid passes in collection sessions. Each returned snap
carries a unique platform-assigned media identifier; the engine checks
this identifier against MongoDB at ingestion time and stores only
previously unseen records, enforcing deduplication via a unique index
on the media identifier. Because the same snap may be returned at
multiple nearby centroids, the stored location is that of whichever
query point retrieves it first. Under concurrent dispatch this is a
race rather than a fixed traversal order, so the assignment of a
boundary snap is not reproducible run to run; its magnitude is bounded
in Section~\ref{sec:limitations}.

Each stored document records the snap's identifier, timestamp
(millisecond epoch), duration, approximate location (the queried grid
centroid, not the user's precise coordinate), title (if present), media
type (video or video-no-sound, populated for ${\sim}$80\% of records),
and media URLs. No user identifiers, account names, engagement metrics,
or social-graph information are collected or stored.

The platform uses a hybrid database design: MongoDB stores the
high-volume, schema-flexible snap metadata, while PostgreSQL handles
structured relational data---user accounts, scraper configurations,
roles, permissions, and group assignments. This separation keeps the
collection path fast (append-only JSON documents) while giving the web
layer the transactional guarantees it needs for user and scraper state.

\subsection{Web Front End}\label{sec:frontend}

The front end is a single-page application built with Next.js that
communicates with the FastAPI back end over REST. Authentication uses
JWT tokens, and the interface adapts to two roles: administrators (who
can manage scrapers and users) and regular users (who can explore and
export data). We describe each functional module below.

\subsubsection{Scraper Management}
\begin{figure}[!t]
  \centering
  \includegraphics[width=\columnwidth]{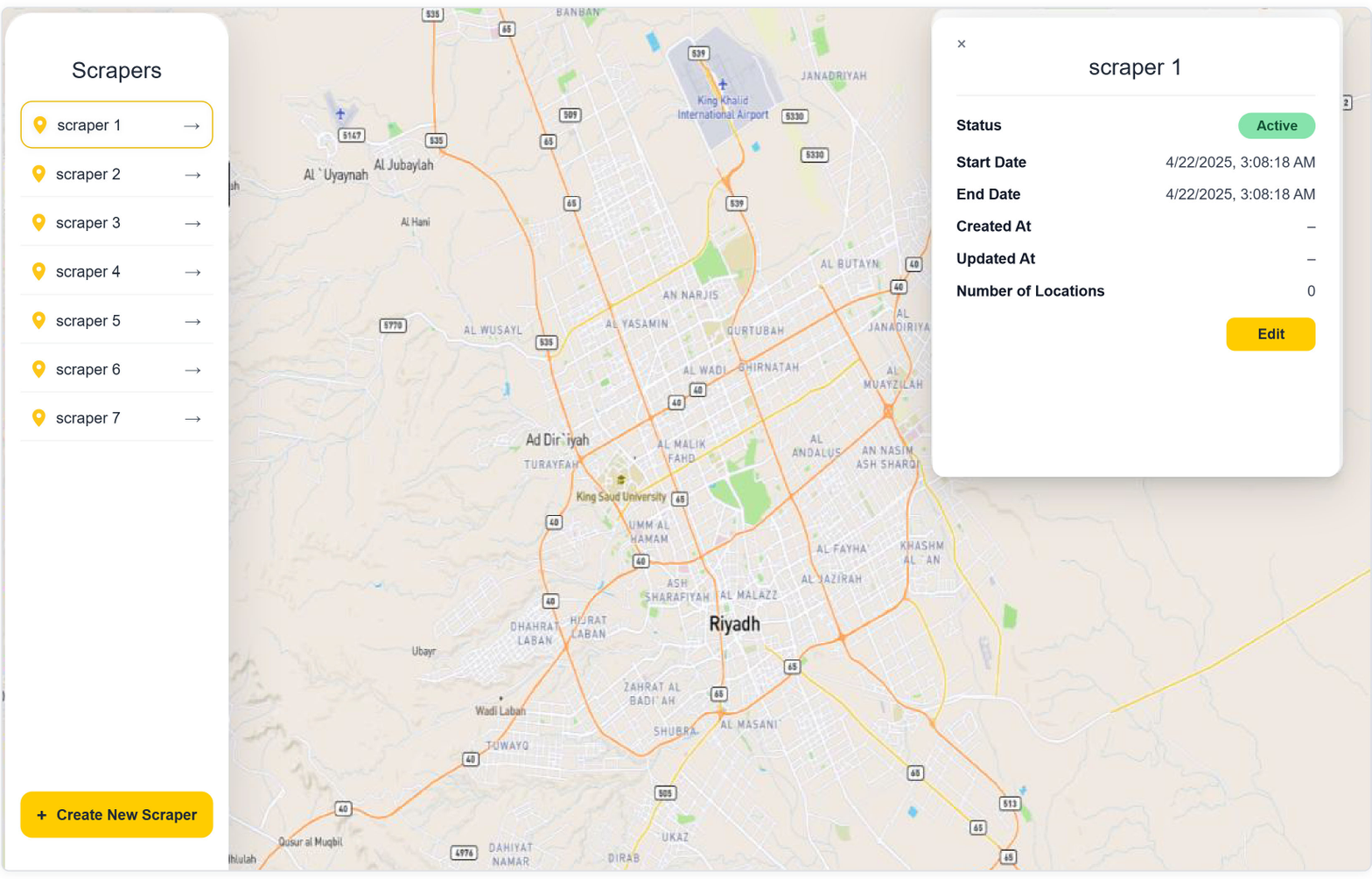}
  \caption{Scraper management interface (demonstration instance).}
  \label{fig:scraper}
\end{figure}

Administrators can create a new scraper by specifying a target city
(selecting or uploading a boundary polygon), a grid resolution, and
start/end dates. Once created, the scraper appears in a sidebar list
with a status indicator (active, stopped, or completed). Clicking a
scraper opens a detail panel showing its progress---start and end
timestamps, current status, and the number of locations and snaps
collected so far. From this panel the operator can stop a running
scraper, edit its parameters, or delete it entirely
(Fig.~\ref{fig:scraper}). During the Riyadh deployment, this interface
was used to monitor pass progress and to restart the scraper.

\subsubsection{Data Exploration and Filtering}
\begin{figure}[!t]
  \centering
  \includegraphics[width=\columnwidth]{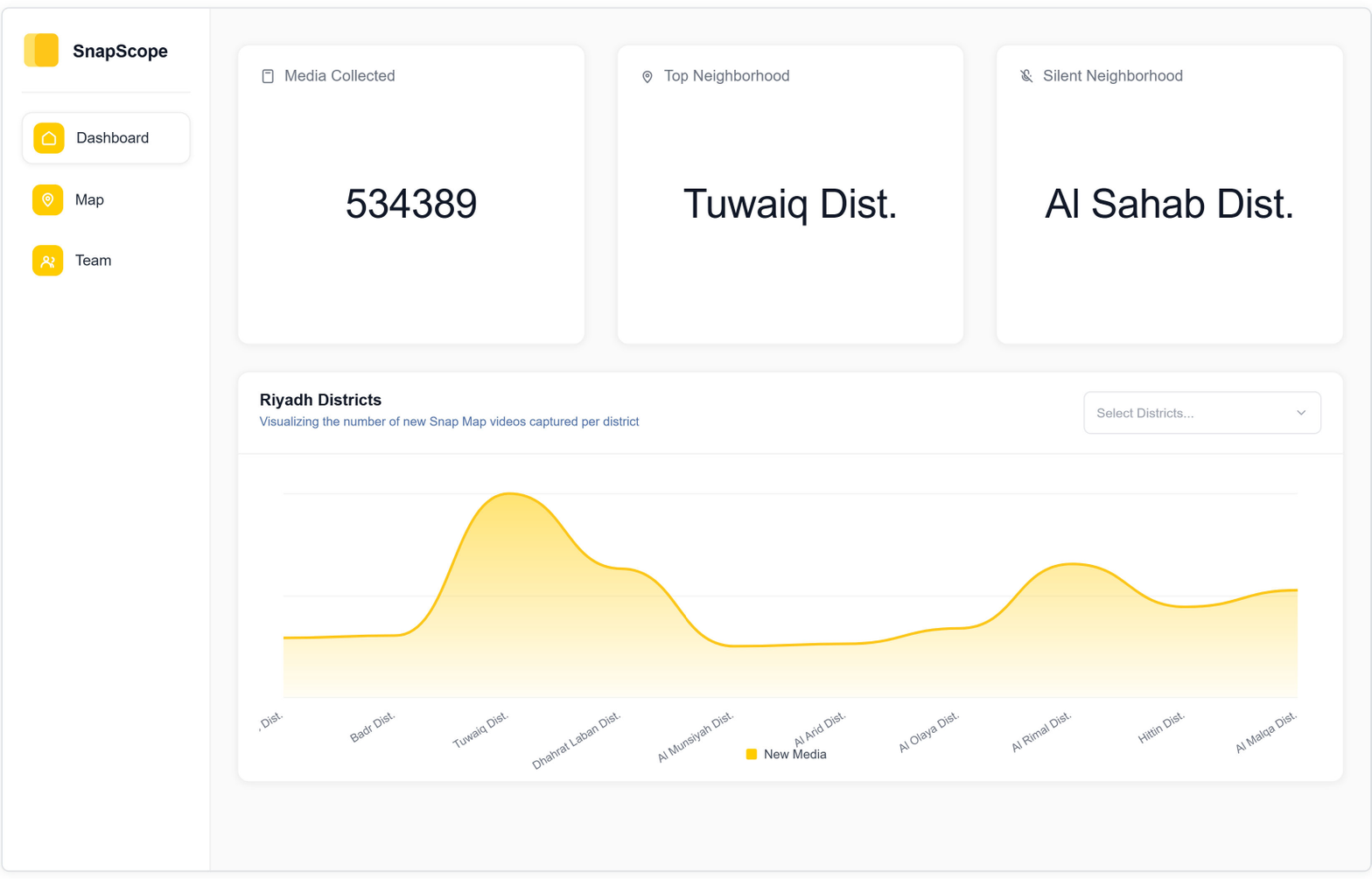}
  \caption{District overview dashboard (demonstration instance), showing the full catalog of 534{,}389~snaps.}
  \label{fig:dashboard}
\end{figure}

The dashboard presents three summary indicators: the total number
of unique snaps collected and the most and least active districts. A
district selector filters a per-district chart of collected media
(Fig.~\ref{fig:dashboard}). A
second dashboard page breaks down per-district media duration and
media type, displays a weekly activity trend line, and shows a pie
chart of district-level shares. All visualizations are built with the
Recharts library and update dynamically as filter selections change.
Users can export filtered subsets as CSV files for offline analysis.

\subsubsection{Advanced Analytics}
A second dashboard page ranks districts by total media duration and
by repeated snap observations detected before deduplication, and
summarizes collection activity by weekday. District filters let users
compare neighborhoods without manual data manipulation or scripting.

\subsubsection{User and Group Management}
The platform implements role-based access control: administrators
manage accounts, assign roles, and organize users into named groups for
collaborative projects, with authentication handled through JWT tokens
issued by the FastAPI back end.

\section{Deployment and Results}\label{sec:results}

\subsection{Riyadh Deployment}

Collection sessions ran over Riyadh on 12 dates from 21~September
through 10~October~2024, covering posts dated 17~September to
9~October. The collection engine ran on a single local machine and
executed complete grid passes in each session, each issuing one API
request per query point across all 2{,}740~points. Requests are
dispatched through a bounded pool of ten concurrent workers rather than
serially, which is what keeps a pass short. Run-level logs are retained
for the 21~passes of the 7--10~October probe window
(Section~\ref{sec:saturation}): a complete pass took a median of
10.6~minutes (mean 11.2, range 9.9--16.9) across the 17~passes that
recorded a completion time, at a cadence of four to six passes per day
and roughly 57{,}500~API requests over the window. Reliability was
imperfect in a way worth reporting: three of the 21~passes recorded no
completion and one returned no data, so roughly one pass in five
required operator attention.

The catalog also holds posts dated outside the window, chiefly from
the days just before it, captured at the first sessions, and from a
final session on 10~October; these are included in the full catalog
(534{,}389~snaps) but excluded from the primary analysis.

Posts from dates without a session were captured from what was still
visible at a later session. Collection paused on 26~September and from
29~September to 5~October; 29~September, with 821~snaps captured, is
treated as incomplete.

\subsection{Dataset Summary}

\begin{figure}[!t]
  \centering
  \includegraphics[width=\columnwidth]{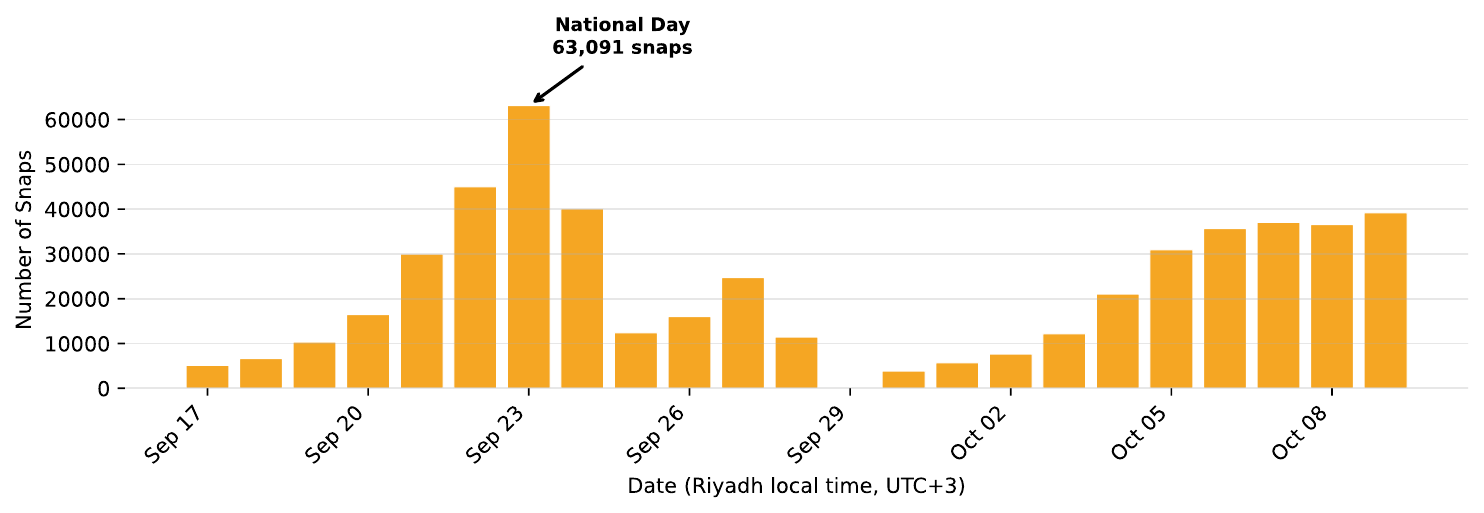}
  \caption{Daily snap volume. The 23~September peak is Saudi National Day; the dip from 29~September reflects the collection pause.}
  \label{fig:daily}
\end{figure}

Table~\ref{tab:dataset} summarizes the collected data.
Figure~\ref{fig:daily} shows the daily snap volume across the 23-day
window. Three features stand out: a sharp peak on 23~September (Saudi
National Day, 63{,}091~snaps in Riyadh local time), a dip on
29~September and lower counts through 5~October, whose posts were
captured from what was still visible when collection resumed. The full catalog---including posts
dated outside the window---contains 534{,}389~snaps; after
filtering to the 23-day primary window the count is 515{,}364, and
after excluding the incomplete 29~September it is 514{,}543.

\begin{table}[!t]
  \centering
  \caption{Riyadh Snap Map dataset summary (17 Sep -- 9 Oct 2024).}
  \label{tab:dataset}
  \small
  \begin{tabular}{lr}
    \toprule
    \textbf{Metric} & \textbf{Value} \\
    \midrule
    Total unique snaps (23-day window)      & 515{,}364 \\
    Snaps in 22 complete days               & 514{,}543 \\
    Full catalog (incl.\ days outside window) & 534{,}389 \\
    Days in window                          & 23 (22 complete) \\
    Grid points queried                     & 2{,}740 \\
    Active tiles (with $\geq$1 snap)        & 2{,}271 \\
    Mean / median snaps per tile            & 226.6 / 139 \\
    Max snaps per tile                      & 2{,}125 \\
    \bottomrule
  \end{tabular}
\end{table}

\subsection{Coverage Saturation}\label{sec:saturation}

To measure coverage stability for Riyadh, we examined a probe window
of 21~consecutive collection runs spanning 7--10~October~2024. Across
these runs the engine encountered 4.8~million snap observations,
of which 94.8\% were duplicates of records already in the database and
5.2\% (248{,}727) were new. This high duplicate fraction indicates
rapid convergence: most content visible on the map at query time had
already been captured in earlier passes.

What this tells us---and what it does not---deserves a clear statement.
The repeated-observation rate shows that successive passes over the
same grid return predominantly already-seen content. It does
\emph{not} tell us what fraction of all public postings the API chose
to surface. Content that expired between passes, or that the API
never returned, appears in neither our collection nor our duplicate
count. The 94.8\% figure therefore measures convergence within the
set the API exposed to us, not coverage of all posting activity in the
city.

\subsection{Spatial Distribution}

\begin{figure}[!t]
  \centering
  \includegraphics[width=\columnwidth]{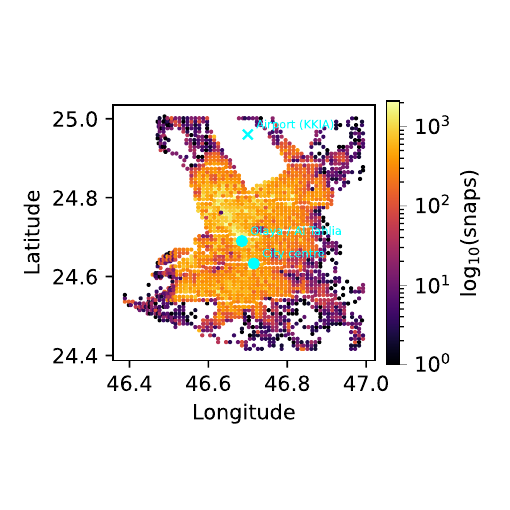}
  \caption{Snap density across Riyadh, 22 complete days (log scale).}
  \label{fig:heatmap}
\end{figure}

Figure~\ref{fig:heatmap} maps the spatial footprint of the collected
snaps. Activity clusters along commercial corridors---particularly
Olaya and Al~Tahlia---and around entertainment zones. The query frame is
built from administrative district polygons, which leave King~Khalid
International Airport outside it; the void there reflects that design
rather than platform behavior. Of the 2{,}271 active tiles in
Table~\ref{tab:dataset}, 2{,}133 resolve to queried grid points once the
140 off-grid coordinates are reassigned (Section~\ref{sec:limitations}),
leaving 607 (22.2\%) of the 2{,}740~queried tiles with zero snaps,
while the top~10\% of active tiles accounted for 36.6\% of the
total. The bottom half contributed only 6.2\%. The distribution is
right-skewed (SD\,=\,275.2 vs.\ mean\,=\,226.6) but broadly spread
rather than funneled into a single downtown core, consistent with
Riyadh's polycentric, car-oriented urban layout. This pattern is
notably less steep than the concentration Juh\'{a}sz and
Hochmair~\cite{juhasz2018} reported for Miami, Los Angeles, and New
York City, where the top~1\% of tiles accounted for nearly a quarter of
all snaps; in our data the top~1\% holds 6.9\%.

\subsection{Implications for Urban Sensing}
The patterns this deployment surfaces are useful because they are
uneven. Activity concentrates along the Olaya and Al~Tahlia corridors
and around entertainment zones, and the top decile of tiles holds
roughly 37\% of all snaps at every resolution we tested---a coarse but
stable indication of where attention gathers, readable without
account-level data. The temporal signal is similarly blunt: the largest
single day is Saudi National Day at 63{,}091~snaps, so a known civic
event registers in raw volume with no event-detection machinery. For a
smart-city operator the feed is therefore a low-latency indicator of
\emph{where} and \emph{when} attention concentrates, useful for directing
more precise instruments rather than as a census of who is present. The ephemerality that makes systematic
collection difficult is the same property that makes the signal
current: the map reports what is drawing attention now.

The 1\,km resolution is a design choice, so we test whether this
concentration depends on it. Aggregating the same snaps to 2, 3, and
4\,km cells leaves the top-decile share essentially unchanged---36.5\%,
37.0\%, and 37.4\% against 36.6\% at 1\,km---indicating that the
concentration is a property of the activity rather than of the lattice.
The top-1\% and bottom-half shares do fall under aggregation (to 3.2\%
and 3.1\% at 4\,km), as expected once the top percentile shrinks from
22~cells to one; this is the modifiable areal unit effect rather than a
change in the underlying distribution. Only coarsening can be tested
after the fact: 1\,km was the collection resolution, so sub-kilometer
sensitivity is not recoverable from these data.

\section{Dataset Release}\label{sec:release}

A privacy-safe aggregate dataset is available under a CC~BY~4.0
license at \url{https://doi.org/10.5281/zenodo.21804079}. The release
contains tile-level hourly snap counts with surrogate tile identifiers,
and no images, videos, precise geographic coordinates, or account-level
data. The identifiers in this release are randomly assigned and carry no
coordinates; companion artifacts archived in the same record use keys
that preserve grid topology, so the privacy argument rests on
aggregation rather than on obscurity. The released quantities are hourly
counts over a 1\,km lattice covering public districts, and identifying a
tile is not identifying a person.

\section{Limitations}\label{sec:limitations}

The collection pipeline depends on undocumented API endpoints that
could change without notice. Our 10-snap round-trip test validates
retrievability and metadata accuracy but not coverage completeness.
The saturation analysis shows convergence within the set the API
returned, but how that set relates to the true universe of public
postings is unknown. Snaps that expired between passes were missed,
creating a bias toward longer-lived content. Several design choices limit
the damage when the endpoint shifts: each request refreshes the map
epoch rather than caching it, requests are retried up to three times
before a point is abandoned for that pass, metadata is stored as
schema-flexible documents so an added or renamed field does not break
ingestion, and deduplication on the platform-assigned media identifier
makes a pass idempotent, so an interrupted run can simply be repeated.
None of this survives withdrawal of the endpoint itself; the pipeline's
longevity is bounded by the platform's willingness to serve it.

Because the stored location is the queried grid centroid rather than
the snap's native coordinate, a snap returned by multiple adjacent
centroids is attributed to whichever centroid retrieves it first, which
may shift individual records across grid-cell boundaries. The
displacement this introduces is small and bounded. Of the 2{,}271
active coordinate tiles, 2{,}131 coincide with a queried grid point and
carry 512{,}141 of the 514{,}543~snaps (99.5\%); the remaining 140 lie
off the grid and are reassigned to their nearest within-district point,
a median of 501\,m away, carrying 2{,}402~snaps between them (0.47\%).
Because 137 of the 140 land on points that are already active, the
active-point count moves only from 2{,}131 to 2{,}133.

The platform observed one city over 23~days of posts; multi-city and
multi-season deployments are needed to test generalizability. The web
front end was used operationally throughout the Riyadh deployment and
passed a structured functional test suite covering scraper creation,
data retrieval, and access control, but functional testing is not
usability evidence: it has not been evaluated through a controlled
study with external users.

\section{Conclusion}\label{sec:conclusion}

We presented SnapScope, an integrated platform for city-scale collection
and interactive exploration of public Snap Map data. The back end
automates the full collection cycle---from API integration through
grid-based querying to ingestion-time deduplication---and the front end
gives researchers direct access to filtering, visualization, comparison,
and export without scripting.

Deploying the platform over Riyadh produced a dataset of
515{,}364~snaps, posted over 23~days, across 2{,}271~active tiles. A saturation analysis
shows that 94.8\% of snap observations across 21~consecutive runs
are duplicates, confirming that successive passes predominantly return
already-captured content. A privacy-safe aggregate dataset is available
under CC~BY~4.0 (\url{https://doi.org/10.5281/zenodo.21804079}).

The platform is city-agnostic: redeployment to a different city requires
only substituting the grid coordinates and boundary polygon. Future work
includes multi-city deployment to test generalizability,
a formal usability study of the web interface, and integration of
automated content classification to add a semantic layer to the spatial
metadata. Together, these extensions would turn a single-city feasibility
demonstration into a general-purpose urban sensing instrument.

\section*{Ethics Statement}
Institutional review board approval is not applicable: this study is
not human-subjects research. It analyzes publicly posted Snap Map
content, collected from the public map interface without logging in,
and the stream carries no username, profile, or account identifier. We
attempted no identification or de-anonymization and report only
aggregate tile-level counts. Downloaded media are retained by the
research team for this and subsequent studies and are not
redistributed.

\section*{Acknowledgment}
The authors used a large language model (Claude by Anthropic) for
language editing and presentation support only, and not to generate
datasets, results, scientific claims, or conclusions. All authors reviewed and approved the scientific
content and take full responsibility for accuracy, originality, and
conclusions.

\IEEEtriggeratref{3}
\bibliographystyle{IEEEtran}
\bibliography{references}

\end{document}